\documentclass[preprint,12pt]{elsarticle}
\usepackage{amsmath,amssymb,etoolbox}

\makeatletter
\patchcmd{\ps@pprintTitle}{\footnotesize\itshape
       Preprint submitted to \ifx\@journal\@empty Elsevier
       \else\@journal\fi\hfill\today}{MITP/14-009}{}{}
\makeatother

\DeclareMathOperator{\quotient}{quo}

\begin{document}

\begin{frontmatter}

\title{A novel approach to integration by parts reduction}

\author{Andreas von Manteuffel and Robert M. Schabinger}

\address{The PRISMA Cluster of Excellence and Mainz Institute of Theoretical Physics, Johannes Gutenberg Universit\"{a}t, 55099 Mainz}

\begin{abstract}
Integration by parts reduction is a standard component of most modern multi-loop calculations in quantum field theory.
We present a novel strategy constructed to overcome the limitations of currently available reduction programs based on Laporta's algorithm. The key idea is to construct algebraic identities from
numerical samples obtained from reductions over finite fields. We expect the method to be highly amenable to parallelization, show a low memory footprint during the reduction step, and allow for significantly better run-times.
\end{abstract}

\end{frontmatter}

Over the past few decades, it has often been the case that new developments in computer technology have sparked advances in theoretical high energy particle physics.
This has been especially true with regard to the application of integration by parts (IBP) identities in $d$ dimensional spacetime to the reduction of multi-loop scalar Feynman integrals
in quantum field theory to a basis of irreducible master integrals \cite{Tkachov:1981wb,Chetyrkin:1981qh}. From the {\tt MINCER} program written long ago for the reduction of three-loop propagator-type integrals \cite{Gorishnii:1989gt},
to the more recent general-purpose algorithm introduced by Laporta \cite{Laporta:2001dd}, automated approaches to integration by parts reduction have long been favored because of the enormous amount of algebra involved.
This is also reflected in the fact that, in recent years, quite a few dedicated IBP solvers have been written and made publicly available \cite{Anastasiou:2004vj, Smirnov:2008iw, Smirnov:2013dia, Studerus:2009ye, vonManteuffel:2012np, Lee:2012cn}.

While many integral reductions of phenomenological interest have been successfully performed in the past,
improvements are required for the calculation of many precision observables relevant to the physics program of the Large Hadron Collider.
For example, solving all of the IBP relations relevant for the calculation of the two-loop virtual corrections to the $p p \rightarrow t \bar{t}$ cross section in Quantum Chromodynamics will take currently available 
reduction programs at least several weeks to run on a desktop computer.
In order to handle future problems, which are likely to be significantly more demanding due to either the presence of additional kinematical scales or additional loop integrations,
it is worth understanding what makes IBP solving computationally expensive.

Let us point out three major performance short-comings of standard
IBP solvers based on Laporta's algorithm.
First of all, for a process on the edge of feasibility, the algorithm will typically require
finding a reduced row echelon form for a sparse system of millions of linear equations with coefficients 
that are polynomial in the available independent ratios of
dimensionful scales and the spacetime dimension.
Solving such linear systems using standard techniques
({\it e.g.}\ variants of Gaussian elimination)
leads to coefficients which are rational functions of high degree
at intermediate stages of the calculation \cite{ModernComputerAlgebra}.
Depending on the exact order of the reduction steps,
the coefficient complexity and the number of nonzero coefficients per
row vector may grow dramatically.

This type of phenomenon is commonly referred to in the literature as
{\it intermediate expression swell}
and leads to performance problems since the expressions become expensive to
manipulate and, en masse, even to store in memory.
For IBP reductions,
a standard operation performed on the coefficients to recognize zeros and to simplify
the resulting expressions is the computation of greatest common divisors,
which becomes increasingly expensive as the coefficients get more and more complicated.
To get a feeling for how severe spurious intermediate expression swell can become
during an IBP reduction, one can mask a single relation between integrals while performing
some set of integral reductions.
Carrying out this experiment, we observed cases where,
as a consequence of the masking, the reduction result grew by more than an order of magnitude in size.
While heuristic rules to avoid expression swell can be found
in available IBP solvers, there is obvious motivation for improvement.

Second, a large fraction of the identities computed in the conventional approach
reduce \emph{auxiliary integrals} which do not occur in the actual calculation
of interest ({\it e.g.}\ some component of a cross section).
However, considering identities involving auxiliary integrals is unavoidable
for a complete reduction of the required integrals.
Clearly, it is of considerable interest to avoid expensive computations
for purely auxiliary quantities whenever possible.

Third, in an effort to improve upon the run-time requirements of the Laporta algorithm, it is natural to attempt a dedicated \emph{parallelization} of the reduction procedure.
Among the publicly available IBP solvers, {\tt Reduze\;2} \cite{vonManteuffel:2012np} is distinguished by the fact that it was designed to be run on a computer cluster.
While the optimal number of cores is problem specific, it is often the case that one observes a significant speed-up only when utilizing up to at most a few tens of cores.
Modern computer clusters available at research institutions and laboratories may provide a considerably larger number of cores which can therefore not be fully exploited.

In this Letter, we describe a new approach to automated integration by parts reduction based on well-known ideas in computational mathematics which
should significantly ameliorate the issues discussed above which one typically encounters in practical applications. Roughly speaking, the strategy is to sample over many distinct prime fields for most of the calculation and
then, at the end, reconstruct the symbolic rational coefficients for the identities of interest by combining the samples together. Remarkably, the requisite mathematical techniques are simple, well-tested, and can be found in expository form
in many modern computer algebra textbooks ({\it e.g.}\ \cite{ModernComputerAlgebra}). Our work is similar in spirit to that of Kant \cite{Kant:2013vta} and, in fact, we expect that his {\tt ICE} package will serve as a useful preprocessor
for IBP relations. The key idea is to systematically avoid manipulating polynomials or rational functions at intermediate stages of the calculation in an effort to avoid intermediate expression swell. 

The outline of this Letter is as follows.
First, we review the reconstruction of rational numbers from samples obtained over
finite fields.
Next, we discuss how this can be exploited for fast rational linear system
solving.
It is possible to work entirely with samples over small (machine-sized) prime
fields, since the information from samples over distinct fields can be combined
by using the well-known Chinese remainder algorithm.
Finally, we promote the rational reconstruction method to the case of univariate rational
functions through interpolating polynomials and discuss various generalizations
and improvements.

Let us begin with a brief review of the mathematical prerequisites. At the heart of everything is the extended Euclidean algorithm (EEA). This algorithm computes the greatest common divisor (GCD) of
two integers, $a$ and $b$, together with their associated {\it B\'{e}zout coefficients}, integers $s$ and $t$ such that
\begin{equation}
{\rm GCD}(a, b) = s\ a + t\ b\,.
\end{equation}
Initially, one begins with the triples $(g_0, s_0, t_0) = (a, 1, 0)$ and $(g_1, s_1, t_1) = (b, 0, 1)$ such that $|a| > |b|$. Then one iterates according to
\begin{align}
q_i &= g_{i-1} \quotient g_i\,\\
g_{i+1} &= g_{i-1} - q_i g_i\,\\
s_{i+1} &= s_{i-1} - q_i s_i\,\\
t_{i+1} &= t_{i-1} - q_i t_i\,,
\end{align}
where $g_{i-1} \quotient g_i$ denotes the integer quotient of $g_{i-1}$ by $g_i$
({\it i.e.}\ $g_{i-1} = g_i q_i + r_i$ for some remainder $r_i$).
The modulus of $g_i$ decreases according to $0\leq \vert g_{i+1} \vert < \vert g_{i} \vert$
until the algorithm terminates with $g_{k+1} = 0$ for some index $k$.
At this point, $g_k = {\rm GCD}(a, b)$, $s_k = s$, and $t_k = t$. It
should be emphasized that the version of the EEA presented above is not guaranteed to be optimal for all integers $a$ and $b$; it will certainly be the case, for example, that a different variant performs better
for $a$ and $b$ with asymptotically large absolute values \cite{ModernComputerAlgebra}.
Throughout this Letter, we will often choose to describe classical versions of algorithms for the sake
of clarity and then point out various optimizations or alternatives which may prove useful.

It turns out that the EEA has a number of useful applications. For example, it is possible to use the EEA to define multiplicative inverses in prime fields, $\mathbb{Z}/p\mathbb{Z}$ (hereafter we use the shorthand $\mathbb{Z}_p$).
If we apply the EEA to $b$ and $p$, we find that
\begin{equation}
1 = s\ p + t\ b
\end{equation}
for some $s$ and $t$. By definition, this implies that \\$1 \equiv t\ b~ {\rm mod}~ p$ and we are therefore led to the definition
\begin{align}
\label{modinvdef}
\frac{1}{b} \equiv t~ {\rm mod}~ p\,.
\end{align}
If we denote the canonical homomorphism from $\mathbb{Z}$ onto $\mathbb{Z}_p$ by $\phi_p(z) = z~ {\rm mod}~ p$, then (\ref{modinvdef}) implies that the {\it $p$-homomorphic image} of a rational number $a/b$ can be consistently written as
\begin{equation}
\label{modratdef}
\phi_p(a/b) = \phi_p(a)\phi_p(1/b)\,.
\end{equation}

The natural question that arises now is whether one can go the other way under certain conditions and reconstruct $a/b$ from its $p$-homomorphic image.
Actually, for our purposes, we must first generalize and replace the prime $p$ with a possibly non-prime positive integer $m$ such that ${\rm GCD}(m, b) = 1$.
Obviously, for the reconstruction to be possible, $m$ must be chosen large enough.
An algorithm to reconstruct $a/b$ from its $m$-homomorphic image was first provided
long ago by Wang \cite{Wang} without proof and then subsequently understood in \cite{WangGuyDavenport}. More recently, this so-called rational reconstruction (RR) algorithm has been improved upon and generalized in a number of important directions
(\cite{Monagan} and \cite{KhodadadMonagan} are of particular interest to us). Before commenting on the state-of-the-art, it is worth saying a few words about how the classical RR algorithm works.

Given two integers $m$ and $u$ fulfilling $u \equiv a/b~{\rm mod}~m$ we want to reconstruct the rational number $a/b$.
The crucial observation is that, when one applies the EEA to $m$ and $u$, one obtains an identity of the form
\begin{equation}
g_i = s_i\ m + t_i\ u
\end{equation}
at every step of the algorithm because the $g_i$, $s_i$, and $t_i$ are computed via exactly the same linear recurrence. Now, if $m$ and the $t_i$ have no common factors, $\phi_m(g_i/t_i) = u$ by definition and it therefore follows that the integers
$g_i$ and $t_i$ obtained at each step of the EEA will {\it all} furnish a rational number, $g_i/t_i$, which is congruent to $u$ modulo $m$.
However, one iteration $j$ turns out to be special and allows one to recover $a/b$
from $g_j/t_j$. Note that, in practice, $m$ will be chosen to be a (relatively large) machine-sized
prime or a product of such primes. This choice for $m$ has the desirable consequence that $m$ and $t_i$ are almost always relatively prime; exceptional cases are very rare and, in any
case, easily dealt with \cite{Kauers}.

We now describe RR as originally envisioned in \cite{Wang}.
Employing the EEA for a generic $m$ as discussed in the previous paragraph,
it can be shown \cite{WangGuyDavenport} that the RR problem will be well-posed
when the modulus $m$ is greater than $2~{\rm max}\{a^2, b^2\}$.
In this situation, the unique solution to the RR problem is given by
\begin{equation}
\frac{a}{b} = \frac{g_j}{t_j}\,,
\end{equation}
where the number $g_j$ is distinguished by the fact that it is the first $g_i$ in the EEA to violate the inequality $|g_i| > \lfloor \sqrt{m/2} \rfloor$. In practical applications, one will usually not know the values $|a|$ and $|b|$
in advance and therefore one needs to veto reconstructions which satisfy either $|t_j| > \lfloor \sqrt{m/2} \rfloor$ or ${\rm GCD}(t_j, g_j) \neq 1$ since, by design, the conditions $|g_j| \leq \lfloor \sqrt{m/2} \rfloor$,
$|t_j| \leq \lfloor \sqrt{m/2} \rfloor$, and ${\rm GCD}(t_j, g_j) = 1$ hold when the RR procedure succeeds.
The point is that, for sufficiently large $m$, all steps of the EEA still yield integers $g_i$ and $t_i$ such that $g_i/t_i \equiv a/b~{\rm mod}~m$ but one step--the $j$th--is special in that both $|g_j| \leq \lfloor \sqrt{m/2} \rfloor$
and $|t_j| \leq \lfloor \sqrt{m/2} \rfloor$.

An important remark is that the RR algorithm outlined above will be suboptimal if $a$ and $b$ are not of roughly equal length.
In fact, it was argued in \cite{Monagan} that, for most practical applications, one must reconstruct rational numbers where  $a$ and $b$ have significantly different lengths.
This indeed appears to be relevant to the problem of interest to us. We note that the modern RR algorithm presented in \cite{Monagan} performs almost as well as the classical variant in unfavorable cases but much better
on average (it succeeds with high probability once $m > 2 |a| |b|$). 

Let us illustrate how the mathematical ideas described
so far can be exploited to construct a fast linear system solver.
Suppose that, for the sake of argument, we want to find a reduced row echelon form for
a large linear system, $A$, with \emph{rational} coefficients.
From the above discussion we see that it is enough to perform a row reduction over a finite
field of size $m$ for sufficiently large $m$. It is then possible to reconstruct
the true rational solution of interest via RR.
Actually, we can go one step further and build up a reduction of $A$
modulo a large number, $m$, with the prime factorization
\begin{equation}
m=p_1 \cdots p_N
\end{equation}
from reductions of $A$ taken modulo the distinct prime factors, $p_i$, of $m$.

For each $p_i$, we take the $p_i$-homomorphic image of $A$ and row reduce $A$
over $\mathbb{Z}_{p_i}$ to obtain a solution set, $K(\mathbb{Z}_{p_i})$.
Here we assume that none of the $p_i$ appear in the
prime factorizations of the denominators of the rational coefficients we wish to reconstruct.
In practice, this condition will be satisfied automatically for large but machine-sized primes
and, at any rate, it is easy to deal with 
exceptions \cite{Kauers}.
Next, we employ the Chinese remainder algorithm (CRA) to produce the solution set
modulo $m$, $K(\mathbb{Z}_{m})$, by sewing together the $K(\mathbb{Z}_{p_i})$ via the EEA. 
As explained above, the solution set $K(\mathbb{Z}_{m})$
can in turn be used to reconstruct the solution with rational coefficients via RR provided $m$ was chosen large enough.
Finally, a check of the purported rational solution, $K(\mathbb{Q})$,
can be performed efficiently working modulo a new prime number.
Employing prime fields smaller than the size of the largest
machine integer has the highly desirable consequence that
fast machine arithmetic can be exploited for the linear algebra.
The scheme described here is well suited for (vector) parallelization and, by the very nature of the CRA, efficiently handles the situation where RR
does not immediately succeed and additional samples must be generated modulo new prime factors before attempting another RR.

We now turn to the problem of finding a reduced row echelon form for a large linear
system whose coefficients are {\it polynomials}.
For the sake of simplicity, we restrict ourselves in this Letter to the case
of a single variable $d$ and consider linear systems with coefficients in $\mathbb{Q}[d]$.
The main non-trivial observation is that a rational \emph{function} of $d$ can be reconstructed from
samples where $d$ is replaced by numbers $\{\alpha_1,\ldots,\alpha_M\}$
for some sufficiently large $M$. In fact, the procedure for rational functions is quite closely analogous to that given above, where solutions over $\mathbb{Q}$ were obtained from samples computed in $\mathbb{Z}_{p_i}$.

Suppose we sample a rational function at the $M$ distinct rational points, $\alpha_r$. Remarkably, writing down the standard Newton interpolation polynomial of degree $M - 1$
which fits the sample data, $I(d)$, furnishes an analog of the CRA in the 
univariate rational function case. Before proceeding, we note that, as for any other polynomial in $\mathbb{Q}[d]$, evaluation at the point $\alpha_r$ is
equivalent to taking the remainder with respect to polynomial division by
$d-\alpha_r$. In other words, we have the simple but useful relation
\begin{equation}
\label{congru}
I(d) \equiv I(\alpha_r)~{\rm mod}~ (d-\alpha_r)\quad\; \forall\; r \in \{1,\dots, M\}\,.
\end{equation}
Now, observe that, by virtue of the fact that univariate polynomial division is completely analogous to integer division, the EEA makes sense
for univariate polynomials as well as integers. In fact, this immediately implies that  Wang's original RR algorithm can be modified to yield a rational
function reconstruction algorithm in the univariate case. One must simply make the observation that the classical RR algorithm is, at its core, nothing more than the EEA with a modified termination criterion.
In particular, Eq. (\ref{congru}) implies that the number $m = p_1\cdots p_{M}$ which appears in the rational number version of Wang's algorithm
is replaced by
\begin{equation}
m(d) = (d - \alpha_1)\cdots (d - \alpha_{M})
\end{equation}
in the univariate rational function version. 

With a bit more analysis (see \cite{ModernComputerAlgebra,KhodadadMonagan}), analogs can be found for all of the other defining characteristics of the
RR algorithm up to essentially trivial differences like uniqueness in the rational function case only holding up to multiplication by a scalar.
In this way, we can reconstruct a rational function for each of the coefficients
in the row echelon form of our linear system from reductions obtained for
rational coefficients. Once rational function reconstruction succeeds for all coefficients in the
row echelon form, we may normalize the vectors obtained and arrive at a solution with entries that lie in $\mathbb{Q}[d]$.

It is worth mentioning that, once again, the classical strategy outlined above can be optimized. For example, the modern variant of rational function reconstruction implemented in the {\tt Mathematica} package\\
{\tt LinearSystemSolver.m} by Kauers \cite{Kauers} was, to the best of our knowledge, first proposed in \cite{KhodadadMonagan}.
The main new insight is that, when one attempts to carry out a rational function reconstruction along the lines discussed above,
almost all steps of the polynomial EEA are such that the rational functions $g_i(d)/t_i(d)$ have total degree $M - 1$ ({\it i.e.}\ the degree of the interpolating polynomial input to the univariate rational function reconstruction algorithm). 
Clearly, if $M$ is large enough to successfully reconstruct the rational function, the $j$th step of the polynomial EEA which actually gives the solution of interest will be such that the total degree of $g_j(d)/t_j(d)$
is less than or equal to the degree of the interpolating polynomial. It follows that a smart strategy is to simply run the polynomial EEA to the end and check whether there is a unique step such that $g_i(d)/t_i(d)$ is of minimal total degree.
If so, it is true with high probability that this step of the polynomial EEA reconstructs the rational function of interest.

Combining the methods discussed so far, we can obtain the reduced row echelon form of a given linear system over $\mathbb{Q}[d]$, $A(d)$,
from prime field samples, $K_r(\mathbb{Z}_{p_i})$, where $r$ indexes the 
sample $\alpha_r$ relevant to the construction of the Newton interpolating polynomial, $I(d)$.
In practice, it is advantageous to avoid introducing rational numbers at intermediate stages of the calculation. For this reason, instead of proceeding exactly as described above,
we reconstruct the row echelon form of $A(d)$, $K(\mathbb{Q}[d])$, according to
\begin{equation}
K_r(\mathbb{Z}_{p_i}) \to K(\mathbb{Z}_{p_i}[d]) \to K(\mathbb{Q}[d])\,.
\end{equation}
In this way, it is possible to construct a fast linear system solver in the
univariate polynomial case which avoids intermediate expression swell and
performs significantly better \cite{Kauers} than traditional linear system
solvers based on some variant of Gaussian elimination.
Although it is not entirely straightforward to treat the multivariate polynomial case,
it is feasible \cite{Kauers,Kant} and will be discussed at length in future work. 

We stress that, in the context of IBP reduction, the approach advocated here allows for
massive (vector) parallelization in the reduction step, since $N\times M$ copies of the same system need to be solved.
Moreover, the reconstruction step is massively parallelizable as well, since, in the IBP relations under consideration,
the rational coefficient function of each master integral may be reconstructed independently of all others.
While existing approaches typically focus on the case of dense systems,
it should be emphasized that IBP systems demand the use of a sparse solver for the linear algebra.
In contrast to standard IBP reduction, we expect the exact order in which the reduction
steps are carried out to have considerably less impact on the performance of the algorithm
as long as sparsity is maintained, since coefficient manipulations have
constant complexity when working over prime fields.

Let us conclude by reiterating that, on general grounds, we expect the IBP solving strategy discussed in this Letter to be considerably more efficient than conventional approaches
based on Laporta's algorithm.
Our method avoids
sources of intermediate expression swell which cause severe problems for
most currently available reduction programs.
It circumvents complicated symbolic manipulations for purely auxiliary
equations and allows for the dedicated reconstruction of the very small subset of IBP
relations actually needed for the specific calculation under consideration.
Finally, it is massively parallelizable and should allow for a much more
effective use of modern computational resources.
\newline
{\em Acknowledgments:}
We would like to thank Manuel Kauers for useful discussions concerning the implementation details of his {\tt Mathematica} package {\tt LinearSystemSolver.m} and Jos\'{e} Zurita for useful comments on a preliminary version
of this work. The research of AvM is supported in part by the
Research Center {\em Elementary Forces and Mathematical Foundations (EMG)} of the
Johannes Gutenberg University of Mainz and by the
German Research Foundation (DFG). The research of RMS is supported by the ERC
Advanced Grant EFT4LHC of the European Research Council, the Cluster of Excellence Precision Physics, Fundamental Interactions and Structure of Matter (PRISMA-EXC 1098).

\end{document}